\documentclass[reprint,pra,citeautoscript,longbibliography]{revtex4-1}

\usepackage[T1]{fontenc}
\usepackage[utf8x]{inputenc}

\usepackage{graphicx}
\usepackage{float}
\usepackage{amsmath}
\usepackage{microtype}
\usepackage{siunitx}
\usepackage{tikz}
\usepackage{pgfplots}
\pgfplotsset{compat=1.14}
\usepgfplotslibrary{fillbetween}

\newcommand{\mytitle}{Adhesive wear mechanisms in the presence of weak
  interfaces:\\ Insights from an amorphous model system}

\usepackage{hyperref} %
\hypersetup{
  final,
  pdfborder={0 0 0},
  pdftitle={\mytitle},
  pdfauthor={Brink, Molinari},
  colorlinks=true,
  urlcolor=blue,
  linkcolor=blue,
  citecolor=blue
}

\usepackage{pdfpages}
\makeatletter
\AtBeginDocument{\let\LS@rot\@undefined}
\makeatother

\makeatletter
\newcommand*{\balancecolsandclearpage}{%
  \close@column@grid
  \cleardoublepage
  \twocolumngrid
}
\makeatother
\begin{document}
\frenchspacing

\title{\mytitle}

\author{Tobias Brink}
\email{tobias.brink@epfl.ch}
\affiliation{Civil Engineering Institute and Institute of Materials
  Science and Engineering, \'Ecole polytechnique f\'ed\'erale de
  Lausanne (EPFL), Station 18, CH-1015 Lausanne, Switzerland}

\author{Jean-François Molinari}
\email{jean-francois.molinari@epfl.ch}
\affiliation{Civil Engineering Institute and Institute of Materials
  Science and Engineering, \'Ecole polytechnique f\'ed\'erale de
  Lausanne (EPFL), Station 18, CH-1015 Lausanne, Switzerland}

\date{May 10, 2019}

\begin{abstract}
  Engineering wear models are generally empirical and lack connections
  to the physical processes of debris generation at the nanoscale to
  microscale. Here, we thus analyze wear particle formation for
  sliding interfaces in dry contact with full and reduced
  adhesion. Depending on the material and interface properties and the
  local slopes of the surfaces, we find that colliding surface
  asperities can either deform plastically, form wear particles, or
  slip along the contact junction surface without significant
  damage. We propose a mechanism map as a function of material
  properties and local geometry, and confirm it using
  quasi-two-dimensional and three-dimensional molecular dynamics and
  finite-element simulations on an amorphous, siliconlike model
  material. The framework developed in the present paper conceptually
  ties the regimes of weak and strong interfacial adhesion together
  and can explain that even unlubricated sliding contacts do not
  necessarily lead to catastrophic wear rates. A salient result of the
  present paper is an analytical expression of a critical length scale,
  which incorporates interface properties and roughness
  parameters. Therefore, our findings provide a theoretical framework
  and a quantitative map to predict deformation mechanisms at
  individual contacts. In particular, contact junctions of sizes above
  the critical length scale contribute to the debris formation.
\end{abstract}

\maketitle

\section{Introduction}

\begin{figure}
  \centering
  \includegraphics[]{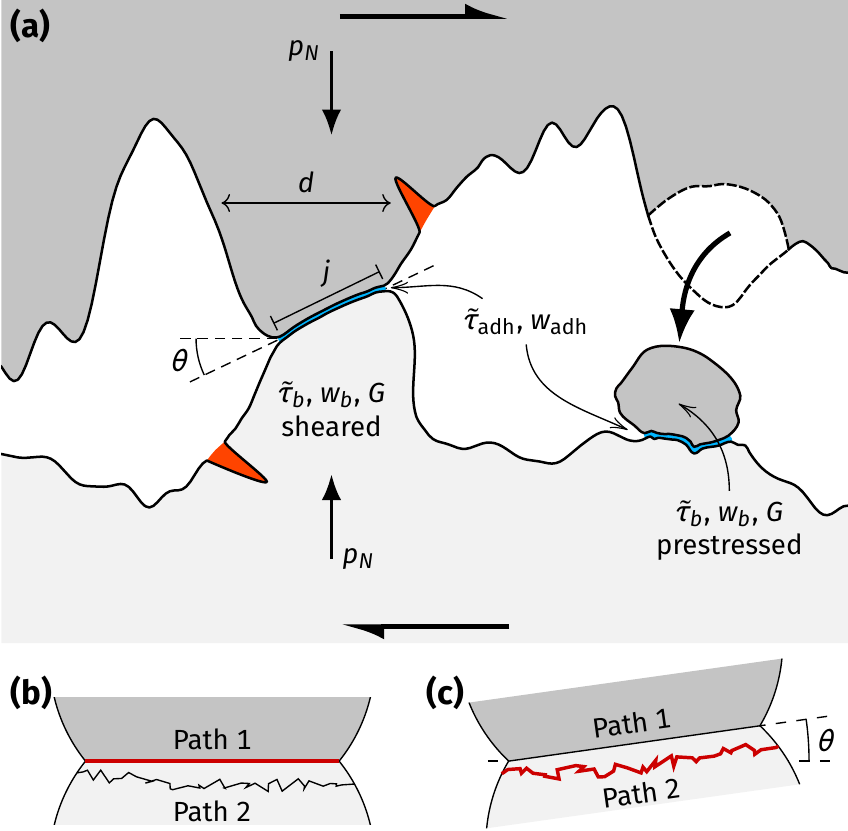}
  \caption{Our current understanding of wear particle creation and
    evolution. (a) Two surface asperities adhere to each other and a
    wear particle starts to detach by cracks (red) propagating in the
    bulk material (left). This process is due to the shearing of the
    contact junction formed between the asperities. A fully formed
    particle might be transferred to the opposite surface and get
    stuck there, based on the strength of adhesion (right). Subsequent
    detachment could be triggered by an additional collision or
    because the particle is prestressed. The roughness of the sketched
    surface is exaggerated compared to typical rough surfaces.  (b) In
    the general case, though, it is still unclear why the cracks
    should propagate in the bulk (path 2) instead of along path 1,
    given that the interface between two surfaces is usually weaker
    than the bulk. (c) One simple, possible solution is to require a
    local slope $\theta$ at the asperity contact, which is in fact
    expected for realistic contacts. Depending on $\theta$, path 2
    could then become the preferred fracture path.}
  \label{fig:intro-sketch}
\end{figure}

\begin{tikzpicture}[remember picture,overlay]
  \node [anchor=north west, font=\footnotesize, align=left,
         text width=7.05in, xshift=0.75in, yshift=0.75in, inner xsep=0pt]
        at (current page.south west)
        {Published in:\\
         \href{https://doi.org/10.1103/PhysRevMaterials.3.053604}
              {T.~Brink and J.-F.~Molinari,
               Phys.~Rev.~Mater.~\textbf{3}, 053604 (2019)}
         \hfill
         DOI: \href{https://doi.org/10.1103/PhysRevMaterials.3.053604}
                   {10.1103/PhysRevMaterials.3.053604}
         \\
         \copyright{} 2019 American Physical Society.};
\end{tikzpicture}%
Contacting solids in relative motion wear, i.e., they gradually lose
volume. Such volume loss is inextricably linked to the roughness and
the details of the contact between the two surfaces \cite{Bowden1939,
  Rabinowicz1951, Burwell1952, Archard1953}. In the case of adhesive
wear, contacting surface asperities form a bond that leads to the
detachment of wear particles during sliding, which are progressively
evacuated from the system if they become loose \cite{Burwell1952,
  Archard1953, Rabinowicz1958}. While such wear particles and the
debris layer they form (collectively called the third body
\cite{Godet1984, Godet1990}) were already observed experimentally a
long time ago \cite{Godet1984}, the process of their creation is still
under discussion. Such an understanding is essential, though, to
establish predictive wear laws \cite{Aghababaei2017, Frerot2018, Molinari2018},
which are currently still limited to empirical formulas that are not
universally applicable \cite{Meng1995}.

Rabinowicz proposed a scenario in which particles become detached from
one of the surfaces and might stick to the other
[Fig.~\ref{fig:intro-sketch}(a) right], possibly growing and becoming
fully loose later on \cite{Rabinowicz1958, Rabinowicz1964,
  Rabinowicz1995}. Similar to Griffith's concept of a critical crack
length for unstable crack propagation \cite{Griffith1921}, one can
define a minimum size a wear particle needs to have before it can
break off by fracture or detach from a surface it is sticking to. This
theoretical argument, though, contains no notion of the asperity
collision process and consequently does not describe a shearing
process as pictured in Fig.~\ref{fig:intro-sketch}(a) left.
Furthermore, it ignores the process proposed by Holm \cite{Holm1967},
in which plastic deformation of the asperities is responsible for
wear. Recently, these theories were combined and extended for the
initial particle formation [Fig.~\ref{fig:intro-sketch}(a) left], by
formulating a critical length scale in terms of a competition between
plastic deformation of an asperity under load and its breaking off
\cite{Aghababaei2016}. In addition to predicting a minimum size of
wear particles, it also provides a way to understand surface roughness
evolution \cite{Milanese2019}: Purely plastic deformation of
asperities has been found to mostly flatten the surfaces or lead to
welding \cite{Sorensen1996, Spijker2011, Pastewka2011, Zhong2013,
  Sha2013, Stoyanov2014, DeBarrosBouchet2015}, while the breaking off
of wear particles seems to be the ingredient to reroughen worked
surfaces \cite{Milanese2019}, although surface kinks due to
dislocation plasticity have also been suggested as sources of
roughness \cite{Venugopalan2019}. The critical length scale is most
generally defined as
\begin{equation}
  \label{eq:d-star-basic}
  d^* = f \dfrac{w_\text{eff}}{\tilde{\tau}^2 / 2G},
\end{equation}
where $w_\text{eff}$ is the effective fracture energy and
$\tilde{\tau}^2 / 2G$ is the elastic energy density of an asperity with
shear modulus $G$ when it is loaded to its elastic shear limit
$\tilde{\tau}$ \cite{Aghababaei2016}. The length $d$ refers to the
diameter of spherical asperities, where asperities with $d < d^*$
deform plastically and those with $d > d^*$ form wear particles. The
prefactor $f$ is a shape factor (see
Appendix~\ref{sec:appendix:shapefactor}), which includes information
about the actual shape of the asperities and the geometry of the crack
that leads to their detachment. The value of the effective fracture
energy depends on the case. For the initial detachment
[Fig.~\ref{fig:intro-sketch}(a) left], two cracks in the material bulk
are needed and thus $w_\text{eff} = w_b$, i.e., the bulk fracture
energy. For the detachment of a sticking particle it is
$w_\text{eff} = w_\text{adh}$, which is the adhesive energy. The
details of $\tilde{\tau}$ will be discussed in the course of the
present paper.

This model, though, neglects the fact that the interface formed
between asperities is usually mechanically weaker than the bulk. This
might be because the junction can effectively be thought of as a grain
boundary, or due to surface contamination and chemical passivation, or
even the presence of lubricants. In such a case, it would seem
reasonable that any fracture occurs along the weaker interface [path 1
in Fig.~\ref{fig:intro-sketch}(b)], as has been pointed out before
\cite{Rabinowicz1995}.  Since that would preclude the formation of
wear particles, several improvements of this simple picture were
proposed: If the junction interface has a well-defined yield stress
$\tilde{\tau}_\text{adh}$, it has been suggested \cite{Aghababaei2016}
that one simply needs to insert it into Eq.~\ref{eq:d-star-basic}
instead of the bulk yield stress $\tilde{\tau}_b$, thereby just
increasing $d^*$.  An alternative solution is to take into account
that the junction interface is inclined with respect to the sliding
direction in most cases \cite{Komvopoulos1985}, which suggests that the
resulting stress distribution might favor fracture in the bulk based on the
inclination angle $\theta$ [Fig.~\ref{fig:intro-sketch}(c)]. Indeed, some
early work on macroscopic junction models supports the latter
hypothesis \cite{Feng1952, Green1955, Greenwood1955}, but is based on
macroscale friction at the interface, which is not applicable on the
relevant atomistic scales at junctions. Furthermore, the possibility
of wear by atom-by-atom attrition has been found in some ultralow
wear conditions \cite{Erdemir2006, Jacobs2010, Bhaskaran2010,
  Scharf2013, Zeng2016, Curry2018}. It seems reasonable to assume that
this process occurs when both wear particle formation and severe
plastic deformation of the surface are suppressed and the surfaces
just slip past each other, but this cannot be explained by the theory
described above, since any adhesive contact in this framework either
leads to wear particle formation or non-negligible
plasticity. Ultralow wear can only take place when both of these
mechanisms are disabled.

In the present paper, we therefore reevaluate the elemental processes
of adhesive wear, by now considering both the strength of the
interface and its local inclination $\theta$, as well as add slip
along the interface to the mechanism map. The competition between
plastic deformation inside the asperities, wear particle formation,
and slip can account for more realistic conditions with reduced wear,
such as lubrication and surface passivation. We develop a simple
stress-based model and test it using molecular dynamics (MD) and the
finite-element method (FEM) in both plane-strain and full
three-dimensional (3D) simulations. In particular, we consider the
question of the nature of interface slip: Is it plasticlike with a
defined yield strength or is it cracklike and activated by stress
concentrations?

\section{Theoretical model}
\label{sec:model}

\begin{figure}[b]
  \centering
  \includegraphics[]{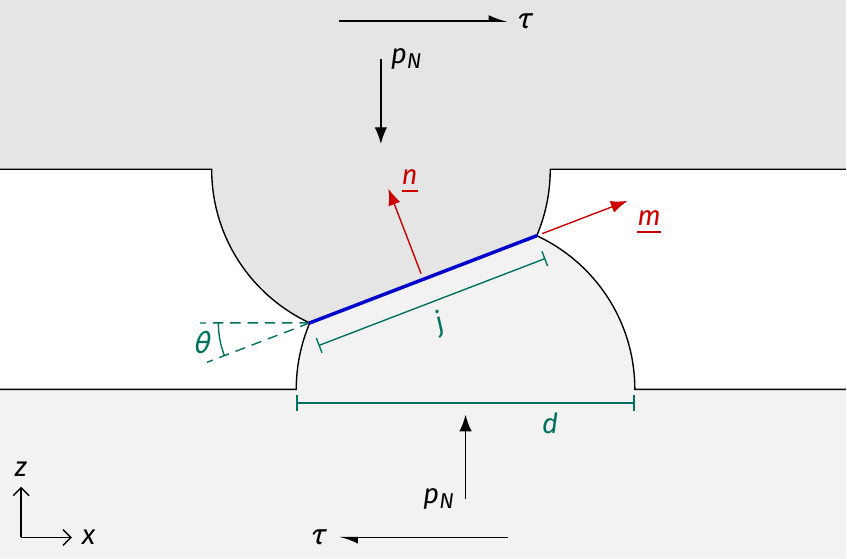}
  \caption{The simplified geometry considered in this paper. The
    contact consists of two overlapping, semicircular (2D) or
    hemispherical (3D) asperities with diameter $d$. The junction
    formed by the overlap has a diameter $j$ and is inclined by an
    angle $\theta$. The interface with normal vector $\underline{n}$
    has a slip direction $\underline{m}$ under the applied shear
    ($\tau$) and normal ($p_N$) loading.}
  \label{fig:geometry}
\end{figure}
We consider the case of the initial detachment of a wear particle
[Fig.~\ref{fig:intro-sketch}(a) left], for which we assume a
simplified geometry of spherical asperities as shown in
Fig.~\ref{fig:geometry}. The basic idea of the critical length scale
model \cite{Aghababaei2016} is that the surfaces keep sliding against
each other, loading the asperities in shear, resulting in a shear
stress $\tau$. At some point either the yield stress of the material
is reached and the asperity deforms plastically, or enough elastic
energy is stored in the asperity to open a crack and detach a wear
particle. This picture needs to be extended, though, in order to be
able to describe slip events at the junction interfaces, which differ
from the plastic deformation of the asperities by causing little
damage in the material, and which might occur in lubricated or
ultralow wear conditions.  Furthermore, as shown in
Fig.~\ref{fig:intro-sketch}(b), it is not clear why fracture in the
bulk should be preferred to slip along the weak interface, especially
since the latter usually has a reduced adhesive energy.

In order to model this, we keep the structure of
Eq.~\ref{eq:d-star-basic}, but will derive a more complete expression
for the critical stress $\tilde{\tau}$. As can be seen in
Fig.~\ref{fig:geometry}, slip events will be limited to the plane with
normal vector $\underline{n}$ in a slip direction $\underline{m}$
perpendicular to it. Thus, we have to consider a resolved shear stress
$\tau_\text{RSS}$ on the interface, similar to a generalized Schmid
factor \cite{Gottstein2004}. We assume that the asperities are loaded
in shear by the sliding and that an additional normal stress $p_N$ is
applied.  This normal stress is generally not equal to the
macroscopically applied load and depends on the actual, local contact
solution. The stress state in the junction is then assumed to be
\begin{equation}
  \label{eq:stress-state}
  \underline{\underline{\sigma}} =
  \begin{bmatrix}
    0    & 0 & \tau \\
    0    & 0 & 0    \\
    \tau & 0 & p_N
  \end{bmatrix}.
\end{equation}
(Note that $p_N$ is compressive for asperities in contact and thus
negative.)  With
\begin{align}
  \label{eq:slip-system}
  \underline{n} &= \begin{pmatrix} -\sin\theta \\0\\ \cos\theta \end{pmatrix} &
  \underline{m} &= \begin{pmatrix} \cos\theta \\0\\ \sin\theta \end{pmatrix},
\end{align}
this results in a resolved shear stress of
\begin{align}
  \tau_\text{RSS} &= (\underline{\underline{\sigma}} \, \underline{n})
                     \cdot \underline{m} \notag\\
  \label{eq:tau-rss}
                  &= \tau\cos2\theta + p_N\dfrac{\sin2\theta}{2}.
\end{align}
A negative $\tau_\text{RSS}$ would mean that the asperities are
driven to slide against the macroscopic sliding direction. In this case the
sliding mechanism is not active because $\tau$ increases with
the sliding distance while the asperities interlock and will thus
increasingly compensate for any ``backsliding'' caused by $p_N$, which
is assumed to be constant. For $\theta \geq 45^\circ$,
the slip mode is thus always disabled and the asperities interlock. Here, only
plastic deformation in the bulk is expected. In fact, the switch from
slippage to interlock occurs at a critical angle
$\theta_c < \ang{45}$, which depends on the normal load and the
material properties.

Of course, in a real contact, the normal vector of the junction is not
constrained to the $xz$ plane. Due to the imposed sliding direction,
though, the slip direction $\underline{m}$ is constrained in this
manner. Hence, the effect on the resolved shear stress is relatively
minor as long as the out-of-plane rotation is lower than \ang{45}, as
shown in Appendix~\ref{sec:appendix:phi}. Higher rotation angles
should rather be interpreted as asperities passing by each other.

\subsection{The crossover angle}

Usually, the strength $\tilde{\tau}_\text{adh}$ of the adhesive
interface is smaller than or equal to the bulk yield strength
$\tilde{\tau}_b$. Sliding will be preferred over bulk plasticity if
$\tau_\text{RSS}(\tau = \tilde{\tau}_b, \theta) >
\tilde{\tau}_\text{adh}$. Thus, if the interface strength equals or
exceeds the bulk strength, sliding is never possible. For all other
cases, by solving Eq.~\ref{eq:tau-rss} with $\tau = \tilde{\tau}_b$
and $\tau_\text{RSS} = \tilde{\tau}_\text{adh}$, we obtain the
crossover angle
\begin{equation}
  \theta_c = \arctan\left(
               \dfrac{p_N/2 \pm \sqrt{\tilde{\tau}_b^2
                                      - \tilde{\tau}_\text{adh}^2
                                      + p_N^2/4}}
                     {\tilde{\tau}_b + \tilde{\tau}_\text{adh}}
             \right).
\end{equation}
Using the ``plus'' sign gives positive angles. We do not discuss
$\theta < 0$ here, since asperity collisions during sliding are
frontal. Such a case might nonetheless occur at the onset of sliding,
or when the sliding direction changes, but then we expect the adhesive
strength to dominate, especially since slip is aided by the normal
load instead of suppressed.

\subsection{A refined critical length scale}

For the moment, we will assume that a well-defined yield strength for
the interface exists. As such, we can retain Eq.~\ref{eq:d-star-basic}
by using a modified critical strength to obtain an angle and
load-dependent critical length scale
\begin{subequations}
\label{eq:d-star}
\begin{gather}
  \label{eq:d-star-repeat}
  d^*(\theta,p_N) = f \dfrac{w_\text{eff}}{\tilde{\tau}(\theta,p_N)^2 / 2G}\text{, with}\\
  \label{eq:tau-crit}
  \tilde{\tau}(\theta,p_N) =
  \begin{cases}
    \tilde{\tau}_b & \theta \geq \theta_c \\
    \dfrac{\tilde{\tau}_\text{adh} - 0.5 p_N\sin2\theta}{\cos2\theta} & \theta < \theta_c
  \end{cases}.
\end{gather}
\end{subequations}
In some special cases $\tilde{\tau}$ is constant and
Eq.~\ref{eq:d-star-basic} is recovered: For full adhesion
($\tilde{\tau}_\text{adh} = \tilde{\tau}_b$, noting that
$p_N \leq 0$), we always obtain $\theta_c = \ang{0}$. This means that
the slip mechanism is disabled and we recover the originally proposed
formula \cite{Aghababaei2016}, where $d^*$ is independent of $\theta$
and $p_N$. The same is true for $\tilde{\tau}_\text{adh} > \tilde{\tau}_b$.
For $\theta = 0^\circ$, i.e., grazing incident, there is also no
dependence on $p_N$ and the bulk shear strength is simply replaced by
the interface shear strength.

\section{Verification using molecular dynamics simulations}

\subsection{Methods and atomistic model}
We used MD simulations in order to test and refine the proposed
model. We chose an amorphous model material, a siliconlike glass,
since worn surfaces often exhibit amorphous top layers
\cite{Scherge2003} and because we want to avoid the complications of
crystal anisotropy for the present paper. Additionally, glasses
plastically deform by shear banding \cite{Talati2009, Fusco2010,
  Falk2011, Greer2013}, which is known to have a well-defined yield
strength that does not vary significantly in the presence of (flat)
surfaces \cite{Cheng2011a, Albe2013}. Dislocation nucleation, on the
other hand, is very sensitive to the nature of interfaces and to size
effects \cite{Gottstein2004}, making an exact yield strength less
accessible for the complex geometry of asperities
\cite{Venugopalan2019}.

To be able to study a large number of geometries, we need to reduce
the computational requirements and thus need a brittle material (small
$d^*$); in our case a siliconlike model material. There is a large
choice of potentials for silicon, but most of them suffer from an
unphysical ductility \cite{Holland1998, Holland1998erratum,
  Pastewka2008, Pastewka2012}. State-of-the-art potentials fixing this
problem exist in the form of screened bond-order potentials
\cite{Pastewka2013}, but in our testing the computational cost is at
least an order of magnitude higher than Stillinger--Weber
\cite{Stillinger1985} or Tersoff-type \cite{Tersoff1988a}
potentials. Instead, we opted for a modified Stillinger--Weber
potential \cite{Holland1998erratum}, in which the three-body parameter
$\lambda_\text{SW}$ (which governs the bond directionality or
stiffness of bond angles) has been doubled from 21 to 42, improving
the description of fracture at the cost of other properties
\cite{Holland1998erratum}. While not suitable to accurately describe
real silicon, the tunable brittleness of this model has proven useful
to study fracture \cite{Holland1998, Holland1998erratum, Hauch1999,
  Holland1999}, as well as the plastic behavior \cite{Fusco2010,
  Fusco2014, Shi2014} and the vibrational properties of glasses
\cite{Beltukov2016}.

All our simulations were performed with \textsc{lammps}
\cite{Plimpton1995}, using a timestep of \SI{1}{fs} for the
integration of the equations of motion. Glass samples were prepared
following the procedure in Ref.~\onlinecite{Fusco2010}: The system was
quenched under full periodic boundary conditions from the melt at
\SI{3500}{K} down to \SI{10}{K} at ambient pressure with a cooling
rate of \SI{e11}{K/s}, after which energy minimization was performed,
yielding a system of size \SI{20x20x20}{nm} with around 400,000
atoms. This initial preparation was performed with the Tersoff
potential for silicon \cite{Tersoff1988a}, since using the
Stillinger--Weber potential during quenching is known to lead to a
glass structure that is too liquidlike in comparison
\cite{Ishimaru1997, Demkowicz2005, Fusco2010}. To obtain the final
sample, the system was equilibrated at \SI{300}{K} and ambient
pressure for \SI{100}{ps} with the modified Stillinger--Weber
potential. The resulting sample has a density of \SI{2.17}{g/cm^3},
Young's modulus of \SI{149}{GPa}, Poisson's ratio of $0.2447$, shear
modulus $G = \SI{60}{GPa}$, and surface energy
$\gamma_s = \SI{1.31}{J/m^2}$.

\begin{figure}
  \centering
  \includegraphics[]{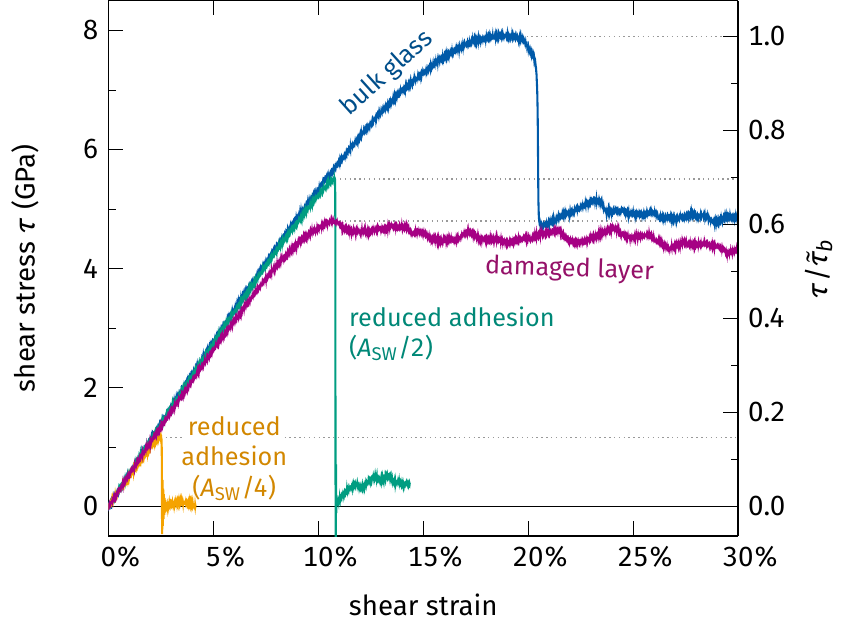}
  \caption{Stress--strain curves of different interfaces under shear
    compared with a homogeneous bulk glass. The axis on the right
    shows the shear stress normalized by the maximum stress
    $\tilde{\tau}_b$ of the bulk glass curve. Snapshots of the
    deformation mechanisms are provided in Fig.~S.1
    \cite{supplemental}.}
  \label{fig:stress-strain}
\end{figure}

For reducing the interface strength, we considered two approaches.
First, we used a modified potential for interactions between the two
sliding bodies, in which we reduced the two-body Stillinger--Weber
parameter $A_\text{SW}$ to half or a quarter of its bulk value. This
leads to a decreased adhesion energy and reduced yield strength under
shear. Second, we randomly removed 20\% of atoms in the interface in a
layer of 1-nm thickness, leading to a disturbed short-range
order as it would occur in a shear band \cite{Ritter2012}. We
investigated the behavior of these interfaces by a shear test on a
bulk sample under periodic boundary conditions with a shear rate of
\SI{4e7}{/s}. Figure~\ref{fig:stress-strain} shows the resulting
stress--strain curves and snapshots are provided in Fig.~S.1
\cite{supplemental}. The bulk material deforms by shear banding, as
expected, connected to the typical, rate-dependent stress overshoot
\cite{Schuh2007}. Reduction of the adhesive energy leads to a complete
loss of shear strength at the onset of sliding, since the energetic
cost of plasticity in the neighboring bulk is higher than for
separation of the interfaces during slip (evidenced by a reduced
coordination number of silicon atoms at the interface and a lack of
plastic events in the bulk), leading to a relatively ``brittle''
stick-to-slip transition. The damaged layer, however, is close in its
behavior to the unmodified bulk material after the onset of plasticity
and shows only a slight softening. Since slip along interfaces can be
thought of as shear banding, these shear curves can be used as an
accurate description of the material behavior in the junctions of our
asperity-contact geometries.

For the sliding simulations, we used the geometry of
Fig.~\ref{fig:geometry} in both quasi-two-dimensional (2D)
plane-strain conditions and in three dimensions, while varying
$\theta$, $d$, $p_N$, and the interface. We kept the junction size
fixed to $j = 0.75d$, except for $\theta \geq \ang{45}$, where we
chose $j = 0.65d$ to avoid overlap between the asperities and the
opposite surface. The geometry was cut out of the bulk glass (repeated
periodically as necessary) in order to allow for an intact short-range
order in the interface. Boundary conditions in $x$ and $y$ were
periodic. We ensured that the bulk part of the surfaces extended at
least a length of $d$ from the asperities in all directions, except
for the largest 3D simulations ($d \geq \SI{50}{nm}$), where we
compromised system size for computational feasibility. The systems
were equilibrated for \SI{100}{ps} at \SI{300}{K} and then a normal
force was applied to a top and bottom layer (width \SI{4}{\angstrom})
of atoms in the $z$ direction to obtain the desired $p_N$ at the junction.
A sliding velocity of \SI{20}{m/s} was chosen and applied at the top
boundary, while the bottom boundary was kept fixed in $x$ and $y$
directions.  Langevin thermostats at \SI{300}{K} with a damping
constant of \SI{0.01}{ps} were applied next to these boundary layers
over a width of \SI{4}{\angstrom}.

\begin{figure*}
  \centering
  \includegraphics[scale=0.95]{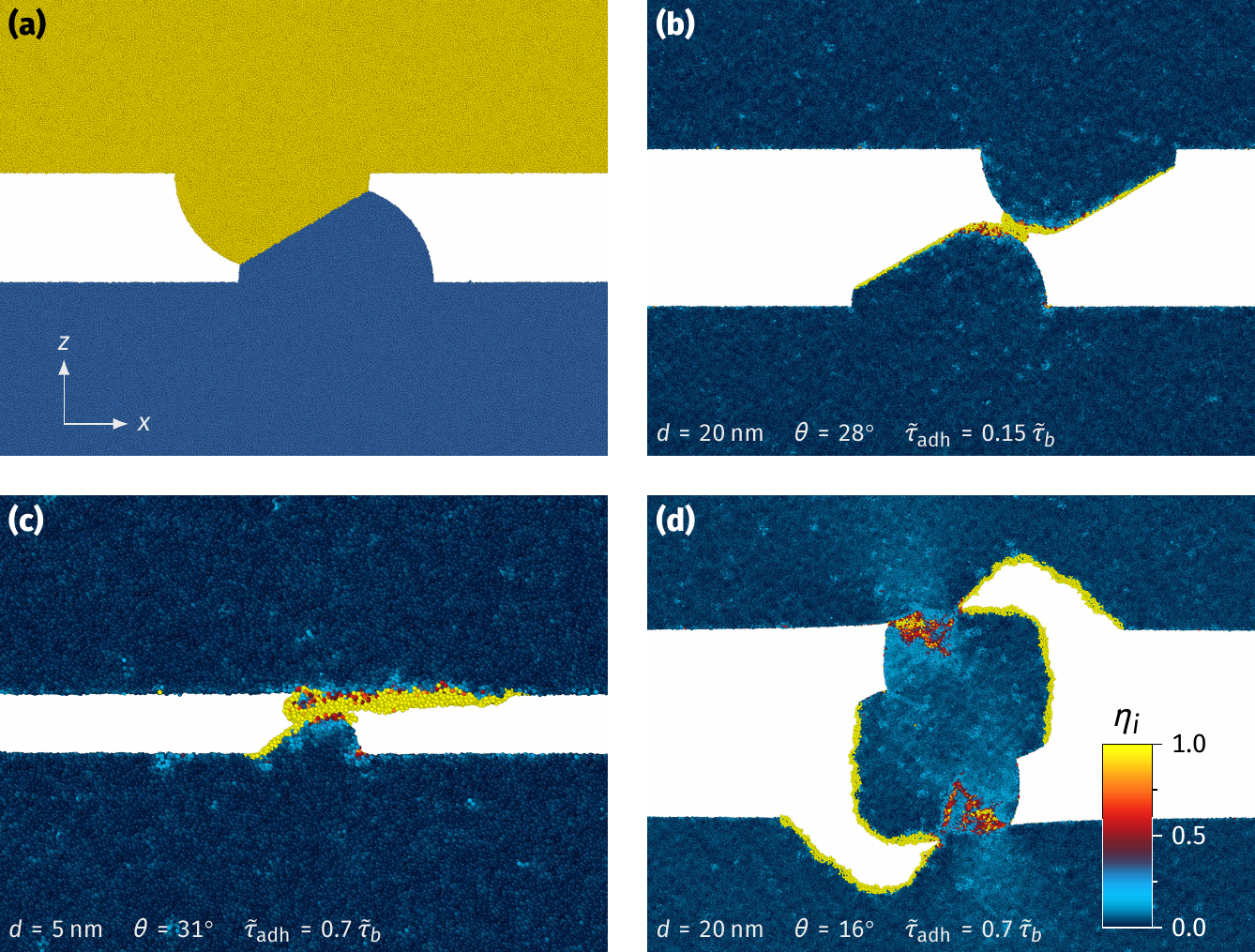}
  \caption{Snapshots of asperity interactions in the quasi-2D,
    plane-strain simulations. (a) An example setup. The colors serve
    to differentiate the two bodies and highlight the interface. The
    snapshots of slip (b), plastic flattening (c), and wear particle
    detachment (d) are colored according to the atomic shear strain
    $\eta_i$. All simulations have a thickness of \SI{20}{nm} in the
    $y$ direction.}
  \label{fig:mechanisms-2d}
\end{figure*}
\begin{figure*}
  \centering
  \includegraphics{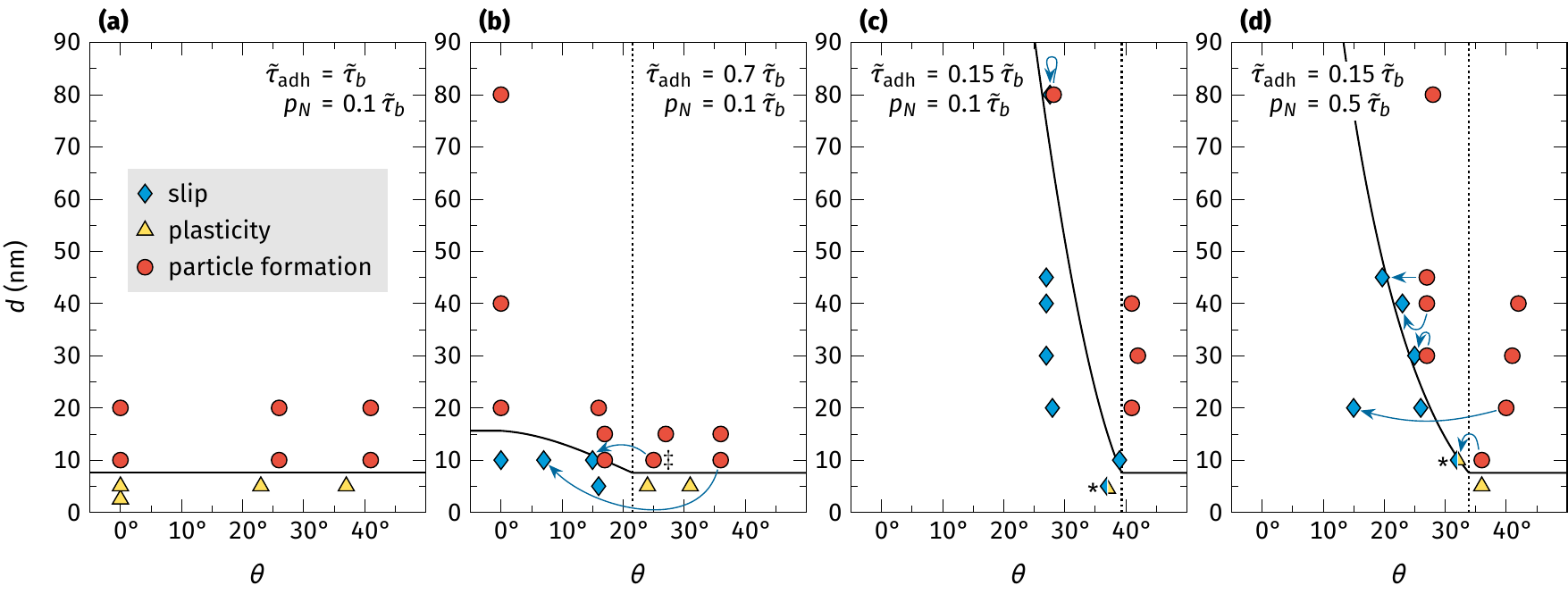}
  \caption{Mechanism map for the quasi-2D simulations. The data points
    represent MD simulations and indicate the observed mechanisms.
    Simulations were conducted with (a) full adhesion at the interface
    between asperities, and (b)--(d) reduced adhesion at the
    interface. (b) Adhesion was reduced to 70\% of the bulk strength
    (green curve in Fig.~\ref{fig:stress-strain}) and (c)--(d) 15\% of
    the bulk strength (yellow curve in Fig.~\ref{fig:stress-strain})
    for two different normal loads. In all figures, the solid lines
    show the prediction for $d^*$ of Eq.~\ref{eq:d-star}: Contacting
    asperities bigger than $d^*$ are expected to form wear particles
    (red). Additionally, the dotted lines indicate $\theta_c$, i.e.,
    the transition from significant bulk plasticity at
    $\theta > \theta_c$ (yellow) to slip at $\theta < \theta_c$ (blue)
    for smaller asperities. The data show that the predictions are
    reproduced well. Some special cases occur, but are in accord with
    the model: Data points connected by arrows indicate a switch from
    an incomplete particle detachment to slip along the
    interface. This can occur due to the rotation of the interface
    while the particle is still forming. An example (simulation marked
    with $\ddagger$) is shown in Fig.~S.2 \cite{supplemental}. The
    simulations marked with an asterisk (*) exhibit a transition from
    slip to plasticity, since the effective normal load $p_N$
    increases during sliding (see Fig.~S.3 in the Supplemental
    Material \cite{supplemental} and the text for a detailed
    discussion). Finally, sometimes slip can be induced by stress
    concentrations close to the junction interface (not shown here,
    see Fig.~S.4 \cite{supplemental}).}
  \label{fig:mech-map-2d}
\end{figure*}

We used \textsc{ovito} \cite{Stukowski2010} for the visualization and analysis
of the simulations. The local von Mises shear invariant $\eta_i$ of
the atomic shear strain was used to identify plastic events in the
glass \cite{Shimizu2007}.

\subsection{Results in plane strain}

In a first step, we performed plane-strain simulations on cylindrical
asperities in a quasi-2D geometry with a thickness of \SI{20}{nm} [see
Fig.~\ref{fig:mechanisms-2d}(a)]. While the critical length scale as
defined by Eq.~\ref{eq:d-star} contains no fit parameters, it was
found in an earlier 2D study \cite{Aghababaei2016} that the shape
factor $f$ is closer to $1.5$ than to the predicted value of
$8/\pi \approx 2.5$. Therefore, we started by performing simulations
with full adhesion for different $\theta$. According to
Eq.~\ref{eq:d-star}, we expect no influence of the contact angle,
since $\tilde{\tau}_\text{adh} = \tilde{\tau}_b$ and thus
$\theta_c = \ang{0}$, meaning that no slip is expected. Indeed, as
plotted in Fig.~\ref{fig:mech-map-2d}(a), we found for all $\theta$
that asperities smaller than $d^* \approx \SI{7}{nm}$ deform
plastically, while larger asperities detach to form wear
particles. Since the fracture of the material is close to perfectly
brittle [see Fig.~\ref{fig:mechanisms-2d}(d)], we assume that
$w_b \approx 2\gamma_s$. Together with the other material parameters
and $d^*$, and by using Eq.~\ref{eq:d-star}, we derive a value of
$f \approx 1.5$, consistent with the earlier findings
\cite{Aghababaei2016}.

Subsequent simulations with reduced adhesion at the interface between
asperities exhibit the three predicted mechanisms of slip, plastic
deformation of the asperities, and detachment of wear particles [see
Fig.~\ref{fig:mechanisms-2d}(b)--(d)]. Slip occurs, as expected,
without significant plasticity inside the two sliding bodies
[Fig.~\ref{fig:mechanisms-2d}(b)]. In the case of plasticity, only one
of the asperities is damaged, while the other stays mostly
intact. This was observed before in simulations of asperity collisions
in diamondlike carbon \cite{Lautz2016} and is a consequence of the
shear softening and strain localization behavior of the material. In
strain-hardening materials, a more even distribution of plastic strain
would be expected.

Figure~\ref{fig:mech-map-2d} shows that the observed mechanisms and
the transitions between them are in accord with the model predictions
of Eq.~\ref{eq:d-star} for different angles, normal loads, and
adhesion potentials. We differentiate between slip and plasticity
inside the asperities by calculating the atomic shear strain $\eta_i$
and checking if plasticity is constrained to the interface or if it
occurs in the bulk. As a criterion, we only consider significant,
irreversible shape changes of the asperities to indicate bulk
plasticity [compare Fig.~\ref{fig:mechanisms-2d}(b) with
Fig.~\ref{fig:mechanisms-2d}(c), for example]. With this, we find that
the critical angle $\theta_c$ is correctly predicted.

Note that in order to accurately compare the results to the model, we
measured the inclination angle $\theta$ of the interface just before
any nonelastic event occurred. We found that $\theta$ can be reduced
by up to \ang{10} during the elastic deformation, although the large
change is a consequence of the large elastic limit of our model
material.

In several cases, we observe that the mechanism can switch, e.g., when
a wear particle starts detaching, but the interface rotates to a
low-angle plane and starts slipping before the crack propagation can
fully unload the contacting asperities (Fig.~S.2
\cite{supplemental}). Nevertheless, the angle at which the switch
occurs is in accord with our theoretical model.
This hints that the detachment of such asperities could be a
fatigue-like process, since the precrack that is formed is not
expected to heal completely and can possibly grow again more easily
upon a following asperity collision.

Moreover, because we control $p_N$ by a constant \emph{force}, the
actual stress at the junction increases during slip. Therefore, we
observe increased plasticity at the end of slip, see
Fig.~\ref{fig:mechanisms-2d}(b) and Fig.~S.3 in the Supplemental
Material \cite{supplemental}. In cases close to $\theta_c$, this
plasticity can be significant. This emphasizes that the evolution of
contacting surfaces is a highly complex and dynamic process, which is
not fully predictable from the initial contact configuration. Our
mechanism map, though, demonstrates that a simplified picture is not
too far from reality and is useful to categorize and understand the
asperity-level mechanisms of wear.

\subsection{Results for the full 3D geometry}
\label{sec:results-3d}

\begin{figure*}
  \centering
  \includegraphics[]{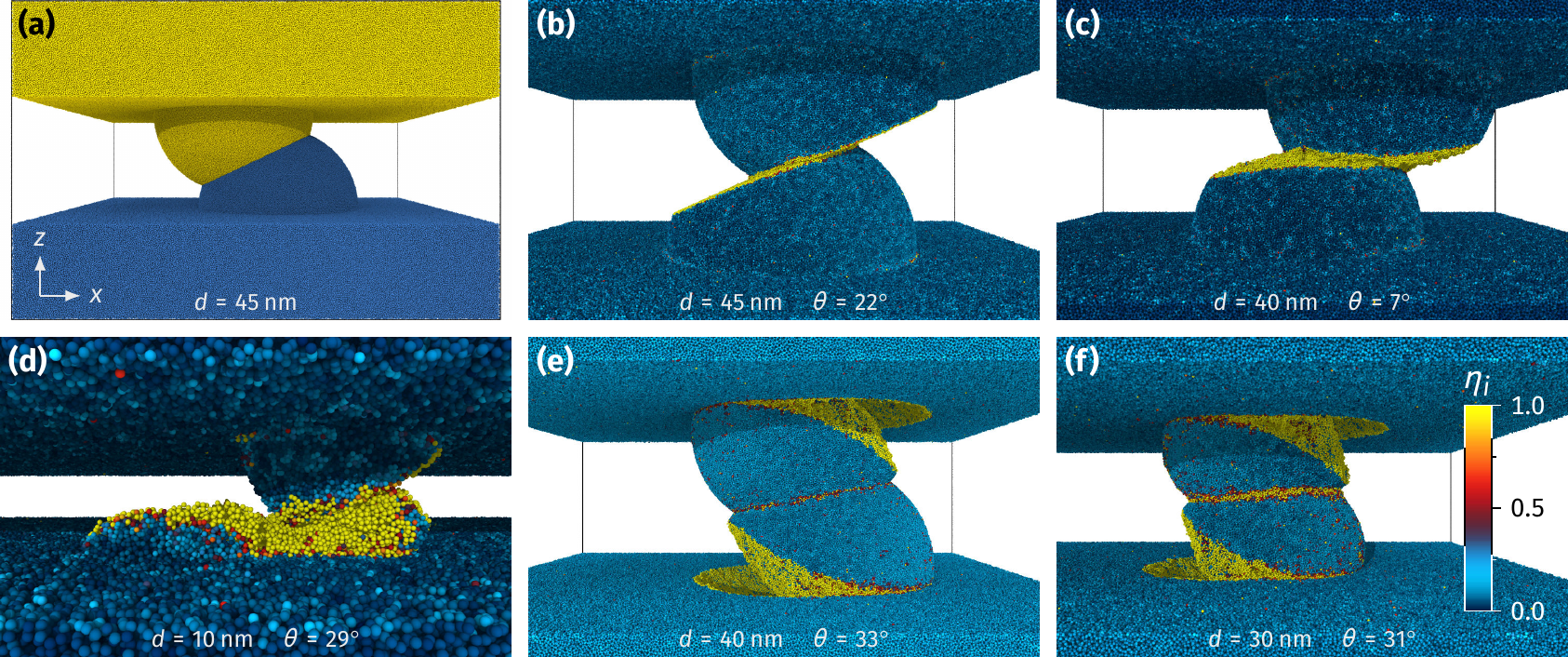}
  \caption{Snapshots of asperity interactions in the 3D
    simulations. (a) An example setup; the colors serve to
    differentiate the two bodies and highlight the interface. The slip
    mechanism is depicted both for the reduced adhesion (b) and the
    damaged (c) interface. (d) Plastic deformation of an asperity with
    reduced adhesion interface. (e)--(f) Formation of wear
    particles. The reduced adhesion interface (e) exhibits less
    plastic activity than the damaged interface (f), but neither
    interface starts slipping.  The snapshots (b)--(f) show zooms on
    the junctions and are colored according to the atomic shear strain
    $\eta_i$.}
  \label{fig:sims-3d}
\end{figure*}

\begin{figure*}
  \centering
  \includegraphics{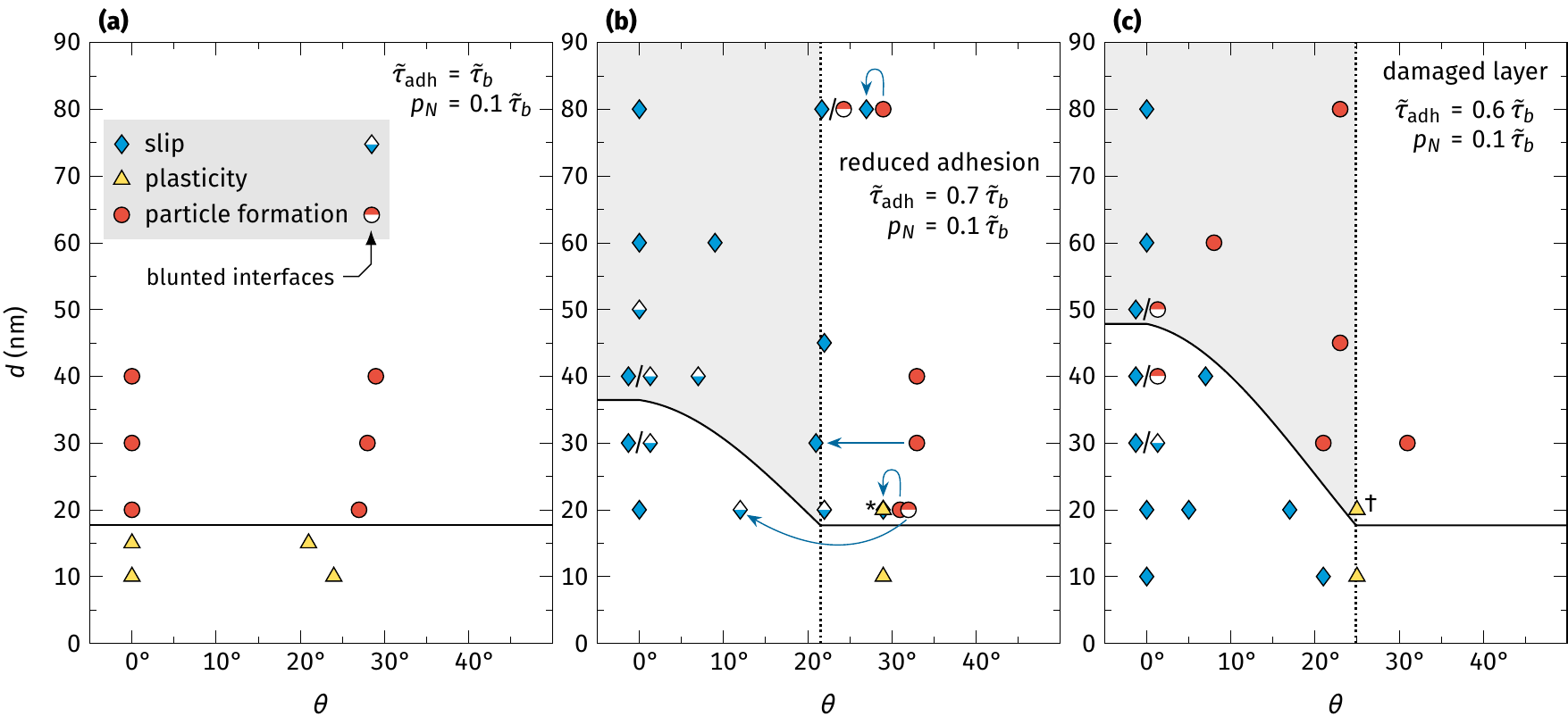}
  \caption{Mechanism map for the 3D simulations. The data points
    represent MD simulations and indicate the observed mechanisms.
    Simulations were conducted with (a) full adhesion at the interface
    between asperities, (b) reduced adhesion (green curve in
    Fig.~\ref{fig:stress-strain}), and (c) a damaged interface layer
    (purple curve in Fig.~\ref{fig:stress-strain}).  In all figures,
    the solid lines show the prediction for $d^*$ of
    Eq.~\ref{eq:d-star}: Contacting asperities bigger than $d^*$ are
    expected to form wear particles (red). Additionally, the dotted
    lines indicate $\theta_c$, i.e., the transition from significant
    bulk plasticity at $\theta > \theta_c$ (yellow) to slip at
    $\theta < \theta_c$ (blue) for smaller asperities. The data show
    that the predictions are reproduced well, except for the gray
    areas in (b) and (c). Here, slip can occur in a regime where wear
    particle formation is expected. For the damaged layer, the
    predicted behavior can be recovered by blunting the interface
    (half-filled data points), while the interface with reduced
    adhesion is not influenced by the blunting. As in
    Fig.~\ref{fig:mech-map-2d}, data points connected by arrows
    indicate a switch from an incomplete particle detachment to slip
    along the interface. This can occur due to the rotation of the
    interface while the particle is still forming. The simulation
    marked with an asterisk (*) exhibited a transition first from
    particle detachment to slip and then from slip to plasticity,
    since the effective normal load $p_N$ increases during sliding
    (cf.\ Fig.~\ref{fig:mech-map-2d}). The simulation marked with a
    dagger ($\dagger$) exhibited a small slip event along the
    interface that triggered a shear band and plastic deformation even
    slightly above $d^*$.}
  \label{fig:mech-map-3d}
\end{figure*}

While the plane-strain simulations are useful to explore the possible
mechanisms of adhesive wear and suggest that our model is predictive,
they do not fully represent all aspects of real contact
geometries. Thus, we extended the simulations to a full 3D geometry by
considering hemispherical asperities [see
Fig.~\ref{fig:sims-3d}(a)]. We again started from simulations with
full adhesion and found $d^* \approx \SI{18}{nm}$ independently of
$\theta$, which, together with the material parameters and
Eq.~\ref{eq:d-star}, corresponds to a shape factor of $f \approx
3.5$. Figure~S.5 \cite{supplemental} shows that the same
results are obtained if the interface is stronger than the bulk, as
predicted. The shape factor of $3.5$ is in fact the expected factor
for overlapping, hemispherical asperities (see
Appendix~\ref{sec:appendix:shapefactor} for the derivation), which
means that Eq.~\ref{eq:d-star} is fully predictive for wear particle
formation in the 3D case under full adhesion conditions without any
fit parameter.

\begin{figure}
  \centering
  \includegraphics[width=\linewidth]{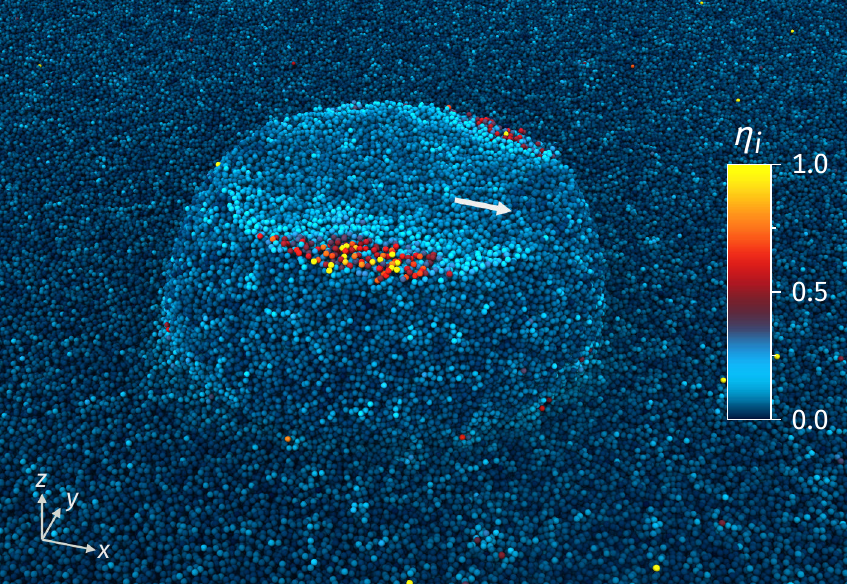}
  \caption{Evidence of a lateral stress concentration in an MD
    simulation, which is responsible for the nucleation of slip. The
    snapshot shows a slice through the junction, with atoms colored
    according to the atomic shear strain $\eta_i$. The arrow on the
    junction indicates the sliding direction, and the onset of slip at
    the lateral corners can be seen.  (The flat surface was made
    darker to make the asperity easier to discern.)}
  \label{fig:stress-conc-md}
\end{figure}

We then extended the simulations to junction interfaces with either
reduced adhesion or a damaged layer. As for the plane-strain
simulations, the three mechanisms of slip
[Fig.~\ref{fig:sims-3d}(b)--(c)], plasticity
[Fig.~\ref{fig:sims-3d}(d)], and particle detachment
[Fig.~\ref{fig:sims-3d}(e)--(f)] were observed. As shown in
Fig.~\ref{fig:mech-map-3d}, the mechanism map partly agrees with the
model prediction, but we unexpectedly found that interfaces with
either reduced adhesion or a damaged interface layer have a tendency
for slip at small $\theta$, even for large asperity sizes $d \gg d^*$
(light gray areas on the graphs). We thus repeated one of the
simulations displaying slip while outputting data in smaller time
increments. Figure~\ref{fig:stress-conc-md} indicates that the start
of slip occurs at edges lateral to the sliding direction, hinting at
slip nucleation by a stress concentration.

In order to elucidate the reason for this difference in behavior
between quasi-2D and full 3D simulations, we turned to finite-element
modelling of our geometry using dynamically inserted cohesive
elements. A detailed description of the setup and results is presented
in Appendix~\ref{sec:appendix:fem}. While this model does not account
for plasticity, it is useful to study the transition from interface
slip to detachment of wear particles. We find that a shear stress
concentration indeed necessarily arises in the same position as in the
MD simulation, which leads to slip along the interface, even for
$d \gg d^*$. Given the existence of a stress concentration, this is
not surprising if the fracture toughness of the junction is reduced
compared to the bulk material: The competition between particle
detachment and slip is not governed by an energetic criterion as in
Eq.~\ref{eq:d-star} anymore, but by a competition between stress
intensity factors. If the junction has full adhesive strength, on the
other hand, the mode I crack at the base of the asperity is always
preferred over the mode II interface slip (see Fig.~S.6
\cite{supplemental}). This also explains why the damaged layer in the
MD simulations has a higher resistance towards interface slip: It does
not weaken catastrophically upon slip and can therefore accommodate
small deformations around the stress concentrator. In other words, it
has a larger plastic zone size and therefore a higher fracture
toughness than the interfaces with reduced adhesion potential (compare
the area under the stress--strain curves in
Fig.~\ref{fig:stress-strain}).

\begin{figure}
  \centering
  \includegraphics[]{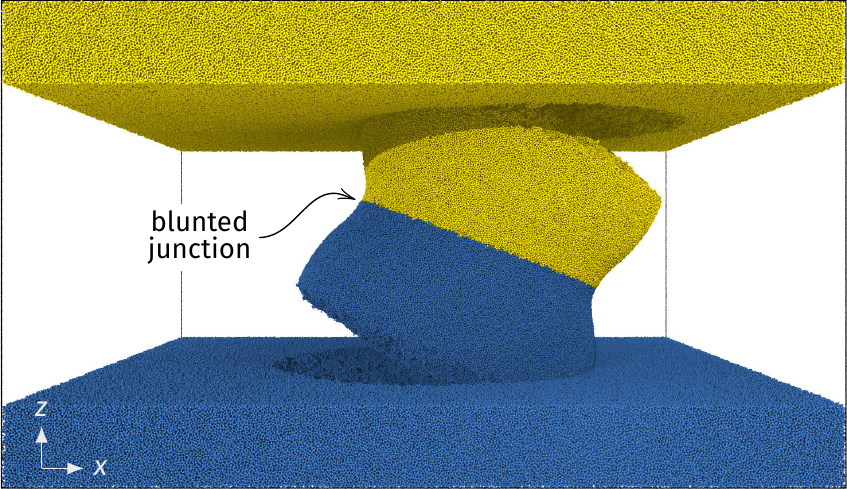}
  \caption{Snapshots of a blunted interface in a 3D simulation with a
    damaged interface layer. The asperity has a diameter of
    $d = \SI{50}{nm} > d^*$ and consequently forms a wear particle as
    predicted by Eq.~\ref{eq:d-star}, in contrast to the same
    simulation with a sharp interface, which exhibits slip along the
    interface [see Fig.~\ref{fig:mech-map-3d}(c) at
    $\theta = \ang{0}$].}
  \label{fig:blunted-3d}
\end{figure}

As such, the competition between slip and wear particle detachment
depends on the details of fracture toughness and relative stress
intensity factors of the two processes. Of course, the simulations
considered until now always possessed sharp interfaces, while contacts
might also exhibit blunted corners at the junction (e.g., due to
diffusion). To explore if we can recover the original model
formulation by reducing the stress concentration, we repeated several
MD simulations with blunted corners around the junction as depicted in
Fig.~\ref{fig:blunted-3d}. We restricted ourselves to simulations
around critical points in the mechanism map in order to reduce the
computational demands. In the case of the reduced adhesion potential
at the interface [Fig.~\ref{fig:mech-map-3d}(b)], no change of
mechanism due to blunting could be observed for $\theta <
\theta_c$. This is most likely because the stress concentration is
only reduced, but not removed by blunting the interface. Due to the
rather abrupt stick-to-slip transition of this interface model, any
onset of slip will be catastrophic. For the damaged layer, though, a
well-defined yield strength and $d^*$ are recovered for all $\theta$
[Fig.~\ref{fig:mech-map-3d}(c)]. The value of $d^*$ for blunted
interfaces with small inclination angles is slightly overestimated. It
should be noted, though, that the damaged interface's strength is
subject to some uncertainty, since the exact damage depends on the
somewhat random local short-range order of the glass.

A switch from wear particle detachment to slip due to rotation of the
slip plane, as observed in the plane-strain case discussed above, is
also present if the interface is ``brittle'' (reduced adhesion), but
not if it is shear-band-like (damaged layer). This emphasizes again
that these processes are ultimately controlled by the exact stress
distributions and process zones that arise. A purely
yield-strength-based argument as laid out in Section~\ref{sec:model}
can only apply if the junction interface behavior is bulklike, which
most likely applies to mature junctions. Additionally, nanoscale
roughness of the junction interface might also help to suppress the
brittle slip behavior. If that does not apply, a competition between a
mode II crack along the interface and a mode I crack that detaches the
particle has to be considered for $\theta < \theta_c$ and $d > d^*$
(gray areas in Fig.~\ref{fig:mech-map-3d}).

Finally, for the derivation of our theoretical model in
Section~\ref{sec:model}, we assumed that a deviation of the sliding
direction by an angle $\phi$ from the $xz$ plane sketched in
Fig.~\ref{fig:geometry} does not change the mechanism map. In
Appendix~\ref{sec:appendix:phi}, we show that this is reasonable for
$|\phi| \leq \ang{45}$ under the assumption that the slip of the
junction will be constrained by the macroscopic sliding direction. In
order to verify this, we repeated the simulations with a damaged
interface, initial angle $\theta = \ang{35}$, and $d = \SI{10}{nm}$,
\SI{20}{nm}, and \SI{30}{nm} and rotated the sliding direction by
$\phi = \ang{30}$. As expected, no change in mechanism was observed, see
Fig.~S.7 in the Supplemental Material \cite{supplemental}. This
means that Eq.~\ref{eq:d-star} is robust even for asperity collisions
that are not perfectly frontal.

\section{Discussion}

The mechanism maps outlined above generally agree with the limited
available literature. Simulations of asperity collisions in
diamondlike carbon were interpreted in terms of the overlap of the
asperities, transitioning from localized shear banding at the contact
spot to a large plastic zone with increasing overlap
\cite{Lautz2016}. Since the asperities were cylindrical and in general
a slightly reduced adhesion due to mismatch of the short-range order
can be assumed, the overlap can be reinterpreted as an angle of
contact, in agreement with our results of transition from slip to bulk
plasticity.

We equally expect the results to be transferable to crystalline
materials, although some complexities have to be taken into
account. Weak planes naturally exist in addition to interfaces in the
form of preferential slip planes and depend on the local lattice
orientation, adding some randomness. Furthermore, while glasses
usually come close to their theoretical strength (if they do not
fracture before) \cite{Cheng2011a, Greer2013}, crystalline plasticity
is size dependent and interfaces with small radii can be stronger than
the bulk \cite{Hurtado1999, Hurtado1999a}. The relevant parameter is
then the shear strength of the weakest part of the asperity,
controlled by the density of preexisting defects.

Beyond the nanoscale, the present paper also has consequences for our
understanding of wear at the engineering scale.
First, large angles between the interface and the sliding direction
lead to an increased probability of wear particle formation, meaning
that a high root-mean-square roughness of the surface leads to more
asperity interlock and more wear.  Consequently, though, the roughness
should reduce during running-in (by wearing off high and thin
asperities), while the real contact area and the size of the
individual contact spots increases and the wear rate
decreases \cite{Blau2005, Milanese2019}. Without the slope-dependent
slip mechanism, this effect cannot be explained, since increased
contact spot sizes should lead to the formation of more and larger wear
particles. Our results provide an answer: The local slopes will be
flatter due to the change of roughness and thus the probability for
slip increases, especially in lubricated conditions.
Second, in a previous study \cite{Frerot2018}, it was found that the
wear coefficient cannot simply be derived only from the assumption
that contact spots larger than a critical value form wear particles,
while smaller ones deform plastically \cite{Aghababaei2016}.
Indeed, this view misses important complexities, such as interactions
between contact spots \cite{Aghababaei2018}, plastic deformations due
to the normal load \cite{Persson2001, Pei2005, Solhjoo2016,
  Venugopalan2019, Frerot2019}, and, as the present paper
indicates, slip.
This slip depends on the strength of the adhesive bond and the
sharpness of the interface, and as such is subject to time-dependent
phenomena such as aging \cite{Li2018c, Li2018b, Dillavou2018} and
interface creep \cite{Coble1963, Ashby1972, Arzt1983,
  Kalcher2017a}. With appropriate knowledge about the parameters of
the interface, wear debris formation, surface deformation due to
plasticity, and slip can now be treated quantitatively using our
framework.

\section{Conclusion}

In conclusion, we revised a criterion for the formation of wear
particles in the adhesive wear regime \cite{Aghababaei2016} by not
only considering the properties of the two bulk materials in contact
but also investigating the interface in more detail. In experiment and
application, weak interfaces between contacting bodies are often
expected and here we find that the interface properties and
orientation play a critical role for wear. For blunted interfaces with
sufficient toughness---as would be expected at junctions that are
sufficiently mature for diffusive processes to have acted---a
well-defined interface shear strength exists. The junction strength in
this case can be derived by a competition between the critical
resolved interface shear stress (comparable to the concept of the
Schmid factor) and the bulk shear strength. Weak interfaces, on the
other hand, are very sensitive to stress concentrators, and will
easily slip in preference to other mechanisms. This means that only
high-angle asperity collisions can lead to wear particle formation.
Finally, when the interface strength approaches the bulk strength, no
slip occurs anymore and only plastic deformation and wear particle
detachment are expected. As a result, flatter surfaces are in general
more amenable to slip and low wear, as long as they are well
lubricated and as long as the contact patches do not grow above a
critical size, which is expected to be very large compared to the
typical contact junction sizes in lubricated conditions. As soon as
lubrication is lost and adhesion increases, the critical size for wear
particle formation will drop drastically and severe adhesive wear is
expected to set in. The present paper provides a theoretical framework
and a quantitative map to predict these transitions.

\vfill

\section*{Acknowledgments}

The authors thank Fabian Barras and Mark O.\ Robbins for helpful
discussions and Nicolas Richart for assistance with the FEM
simulations.
This work was supported by a grant from the Swiss National
Supercomputing Centre (CSCS) under project ID s784, as well as by
{\'E}cole polytechnique f\'ed\'erale de Lausanne (EPFL) through the
use of the facilities of its Scientific IT and Application Support
Center.

\appendix

\section{Shape factor}
\label{sec:appendix:shapefactor}

The shape factor $f$ in Eqs.~\ref{eq:d-star-basic} and \ref{eq:d-star}
contains information about the actual shape of the asperities and the
geometry of the cracks needed to detach a wear particle. In the case
of a contact between two asperities, each of volume $V$, the maximal
stored elastic energy is $2V \tilde{\tau}^2 / 2G$. Two cracks with
area $A$ are needed to form a particle, which requires a work of
$2Aw_b$. Often, the assumption of hemispherical asperities with
diameter $d$ and with a circular fracture area at their base is made
\cite{Rabinowicz1958, Rabinowicz1964, Aghababaei2016}: $A = \pi d^2/4$
and $V = \pi d^3/12$. The balance of the two energies gives
$d^* = 3 w_b/(\tilde{\tau}^2/2G)$ and thus $f = 3$, as shown in
earlier work \footnote{Please note that in the original publication
  \cite{Aghababaei2016}, the factor 2 in front of $G$ was moved into
  $f$ and $w$ included a factor 2 to account for the fact that two
  cracks are needed for particle formation. These differences to our
  formulation cancel out and we arrive at the same value of $f$.}.

In our simulations, the asperities are no longer perfectly
hemispherical. Thus, in order to derive the exact shape factor $f$ for
our 3D geometries, we first consider the volume of the
asperities. Since they are overlapping hemispheres, each asperity can
be described as a spherical segment (see Fig.~\ref{fig:geometry}). A
spherical segment is a solid bounded by two parallel planes cutting
through a sphere \footnote{Note that while asperities with
  $\theta \neq \ang{0}$ are not spherical segments, they have the same
  volume as long as the plane for cutting the junction is rotated
  around the sphere's center and as long as the cuts do not
  intersect. Both conditions are true for all geometries we used.},
resulting in two circular surfaces of radii $a$ and $b$ separated by a
height $h$, and a volume of $\pi h (3a^2 + 3b^2 + h^2) / 6$. Since one
cut goes through the sphere's center, $a = d/2$. The second cut
represents the junction, for which we chose a radius of
$b = j/2 = 3d/8$.  Together with the relation
$d^2/4 = a^2 + ((a^2 - b^2 - h^2)/2h)^2$, we obtain a volume of
\begin{equation}
  V = \frac{41\sqrt{7}\pi}{1536} d^3
\end{equation}
per asperity. We keep the assumption of a relatively flat crack path,
corresponding to the circular base of the asperities. As above, we
solve for the critical length by equating the potential energy with
the work for crack propagation and obtain
\begin{equation}
  d^* = \frac{384 \sqrt{7}}{287} \frac{w_b}{\tilde{\tau}^2 / 2G}
        \approx 3.54 \frac{w_b}{\tilde{\tau}^2 / 2G}.
\end{equation}
The result of $f \approx 3.54$ agrees with the data in
Section~\ref{sec:results-3d} without a fit parameter.

Finally, one can also attempt to define a critical size for a
hemispherical wear particle sticking to one of the surfaces. Using the
same arguments as before, we obtain a maximal stored elastic energy of
$V \tilde{\tau}^2 / 2G$ and a total adhesive energy of
$A w_\text{adh}$, also resulting in $f = 3$. This mechanism, though,
is not as well defined: Rabinowicz assumed that the detachment is due
to residual \emph{compressive} stress and thus used yield strength in
tension and Young's modulus instead of shear strength and shear
modulus \cite{Rabinowicz1958}. But in this case the residual tensile
stress is most likely smaller than the yield strength. Rabinowicz thus
made a back-of-the-envelope estimation that the residual stress is
around 10\% of the yield strength and thereby arrived at $f = 30$, but
did not justify this choice any further. Therefore, this mechanism is
not yet fully understood, neither quantitatively nor qualitatively.

\begin{figure}
  \includegraphics[]{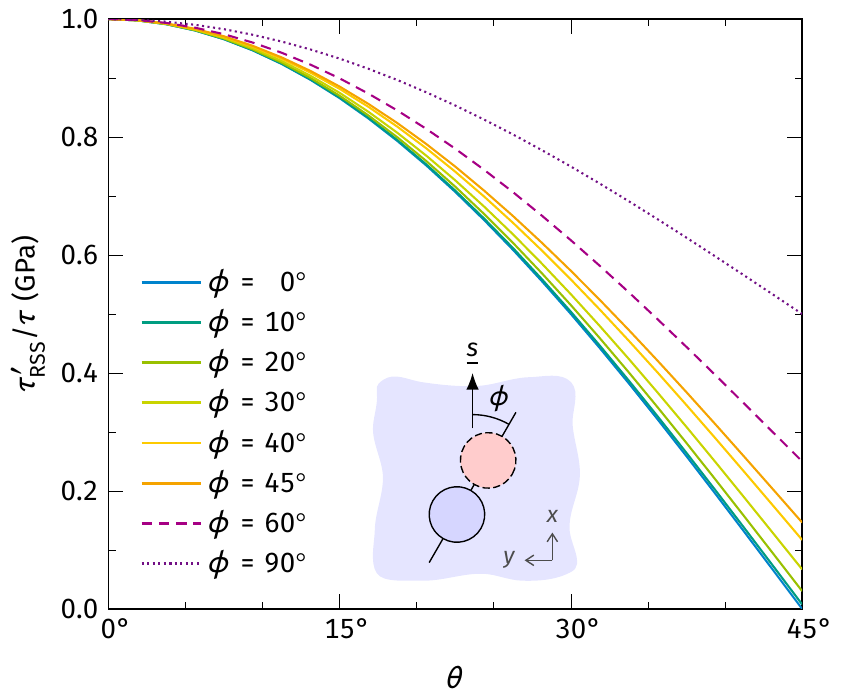}
  \caption{Influence of out-of-plane rotation $\phi$. With increasing
    $\phi$, the resolved shear stress in the junction interface
    increases, thereby promoting interface slip. For
    $\phi \leq \ang{45}$ and low $\theta$, though, this increase is
    very small.  Values are plotted for $p_N = 0$, since the influence
    of the normal load is independent of $\phi$. The inset shows the
    definition of $\phi$, sketched as a top view of an asperity
    contact: The blue asperity is connected to the blue bottom
    surface, sliding in direction $\underline{s}$, while the red
    asperity belongs to the top surface (not pictured), which slides
    in direction $-\underline{s}$.}
  \label{fig:influence-of-phi}
\end{figure}

\section{Rotation of the junction interface normal vector
  out of the $xz$ plane}
\label{sec:appendix:phi}

In a 3D geometry, the derivation of resolved shear stress
(Eqs.~\ref{eq:stress-state}--\ref{eq:tau-rss}) is only valid if the
normal vector of the slip plane remains in the $xz$ plane. We will
show below that a rotation out of this plane by an angle $\phi$ does
not significantly change the resolved stress, as long as the slip
direction remains in the $xz$ plane. The latter is a reasonable assumption,
since the sliding direction is macroscopically imposed.

\begin{figure*}
  \centering
  \includegraphics[]{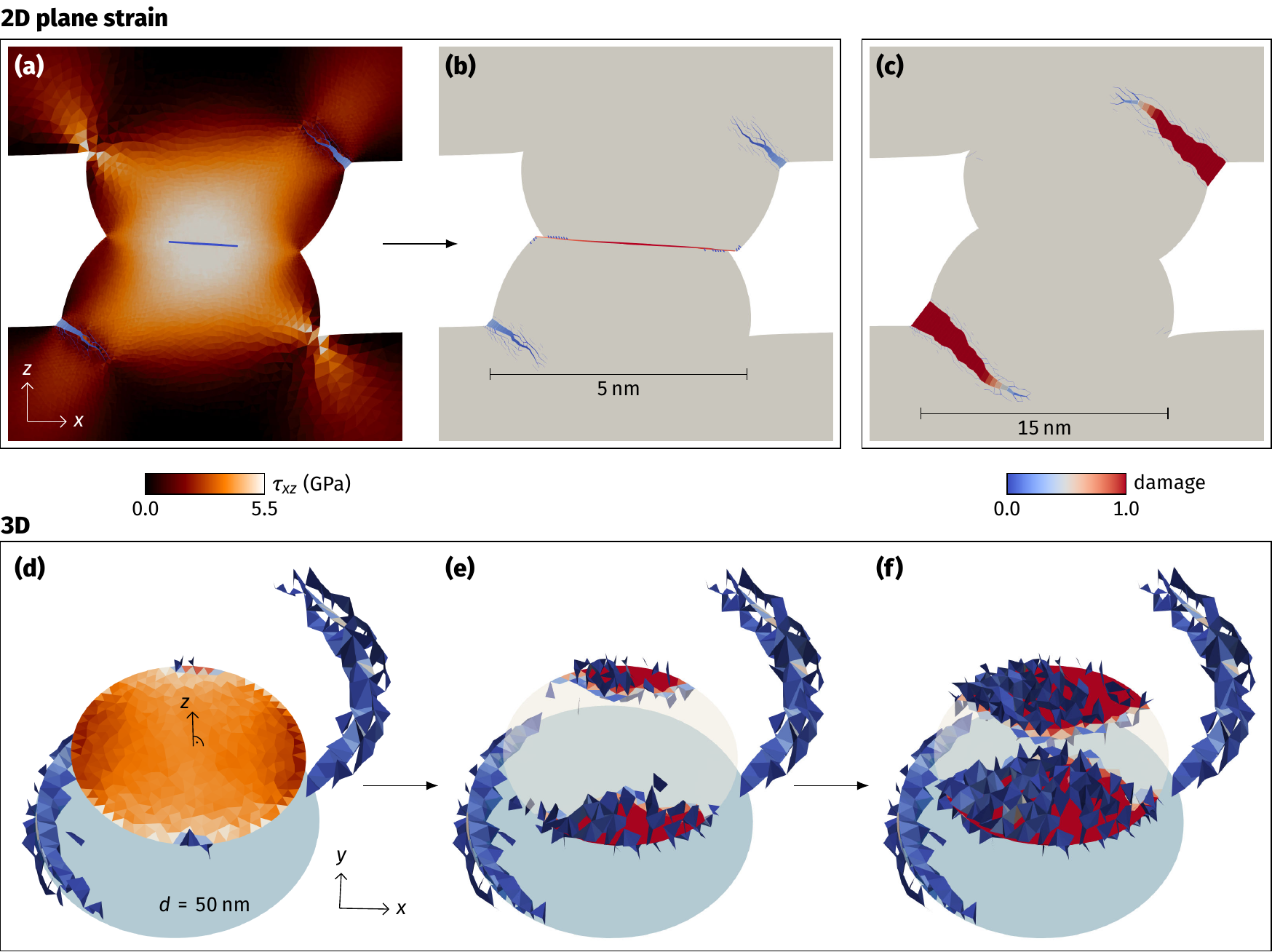}
  \caption{FEM analysis of the difference of stress states between the
    plane strain and the 3D geometries for junctions with reduced
    adhesion. (a) In the 2D case of an asperity with $d < d^*$, the
    interface starts slipping in the junction center, where the shear
    stress $\tau_{xz}$ is highest. The bulk is colored according to
    the shear stress, while the cohesive elements are colored
    according to the damage parameter. (b) After the onset of slip,
    the cracks at the base do not open and wear particle formation is
    suppressed. (c) For $d > d^*$, no slip events can be observed and
    crack propagation leads to particle formation. (d) For the 3D
    case, only cohesive elements and a slice of the junction are
    shown. The blue circle represents the base of the bottom asperity,
    while the junction is colored according to the shear
    stress. Stress concentrations lateral to the sliding direction $x$
    are visible and act as nucleation sites for slip (e)--(f). In
    contrast to the predictions of Eq.~\ref{eq:d-star}, such slip
    occurs even for sizes $d \gg d^*$.}
  \label{fig:fem-2d-3d}
\end{figure*}

The additional rotation can be included in the normal vector
$\underline{n}$:
\begin{align}
  \label{eq:slip-system-phi}
  \underline{n}' &= \begin{pmatrix}
    -\sin\theta \cos\phi \\+\sin\theta \sin\phi \\ \cos\theta
  \end{pmatrix} &
  \underline{m} &= \begin{pmatrix} \cos\theta \\0\\ \sin\theta \end{pmatrix},
\end{align}
resulting in a resolved shear stress of
\begin{align}
  \tau_\text{RSS}' &= (\underline{\underline{\sigma}} \, \underline{n}')
                      \cdot \underline{m} \notag\\
  \label{eq:tau-rss-phi}
                  &= \tau (\cos^2\theta - \sin^2\theta\cos\phi)
                     + p_N\dfrac{\sin2\theta}{2}.
\end{align}
The term relating to $p_N$ is unchanged.
Figure~\ref{fig:influence-of-phi} shows that the change of
$\tau_\text{RSS}'$ with $\phi \leq \ang{45}$ is minor and can mostly
be ignored. As discussed in Section~\ref{sec:results-3d} and shown in
Fig.~S.7 \cite{supplemental}, this is confirmed by simulations with
$\phi = \ang{30}$, which exhibit the same mechanisms as the
simulations with $\phi = \ang{0}$. For $\phi > \ang{45}$, slip is
preferred, which should be interpreted as asperities passing by each
other.

\section{FEM simulations of slip at the junction}
\label{sec:appendix:fem}

In order to elucidate the reason for the difference in the
slip-to-fracture transition between quasi-2D and full 3D simulations,
we turned to simplified finite-element modeling of our geometry. We
used the software \textsc{akantu} \cite{Richart2015, Vocialta2017} to model a
homogeneous, isotropic, linear elastic material with large
deformations. Fracture was modeled using cohesive elements with the
linear irreversible law by Snozzi and Molinari \cite{Snozzi2013}. The
simulations were performed with an explicit time-integration scheme,
where the cohesive elements were inserted dynamically
\cite{Vocialta2017}. The geometry and the sliding velocity of
\SI{20}{m/s} match the MD simulations at $\theta = \ang{0}$.
Visualization was performed with \textsc{paraview} \cite{Ahrens2005}.

We chose parameters to match the model glass and the interface with
reduced adhesion ($\tilde{\tau}_\text{adh} = 0.7\tilde{\tau}_b$) in
the MD simulations: For the linear elastic, isotropic bulk material, a
density of \SI{2.17}{g/cm^3}, a Young's modulus of \SI{149}{GPa}, and
a Poisson's ratio of $0.2447$ were used.

The cohesive elements inserted in the bulk of the material had a
critical insertion stress $\sigma_c = \SI{7.9}{GPa}$ (the shear yield
strength of the glass) with a critical effective opening displacement
of $\delta_c = \SI{0.66445}{nm}$ at which cohesion is lost. This
corresponds to a fracture energy of \SI{2.62}{J/m^2}, which was the
same for mode I and mode II opening (see Ref.~\onlinecite{Snozzi2013}
for details). For cohesive elements inserted along the junction
interface, we reduced the toughness by setting
$\sigma_c = \SI{5.5}{GPa}$ and $\delta_c = \SI{0.2}{nm}$ for a
fracture energy of \SI{0.55}{J/m^2}.

While this model cannot account for plasticity, it is useful to study
the transition from interface slip to detachment of wear particles.

Figures~\ref{fig:fem-2d-3d}(a) and (b) show a 2D plane-strain
simulation with $d < d^*$. Clearly, the shear stress is maximal in the
center of the contact junction, where a mode II crack nucleates and
propagates, leading to slip. The cracks at the bottom of the asperity
never fully form (damage much smaller than $1$). For $d > d^*$
[Fig.~\ref{fig:fem-2d-3d}(c)], on the other hand, the interface never
damages and a wear particle is detached.

The 3D case, though, shows a deviation from our initial
assumptions. In Fig.~\ref{fig:fem-2d-3d}(d), a shear stress
concentration is visible at the junction, lateral to the sliding
direction. Despite $d \gg d^*$, interface slip starts at the stress
concentration and propagates inwards. This matches the MD simulation
with high time resolution shown in Fig.~\ref{fig:stress-conc-md}. When
the modeled junction interface has the same properties as the bulk
material, on the other hand, a wear particle detaches by fracture at
the base of the asperities (see Fig.~S.6 in the Supplemental Material
\cite{supplemental}).

An analysis of the stress state in the junction shows the principal
difference between the 2D and 3D case: In the $xz$ plane, the free
surfaces of the asperities require that
\begin{align}
  \label{eq:boundary-xz}
  \small
  \begin{bmatrix}
    \sigma_{xx} & \tau_{xy} & \tau_{xz} \\
    \tau_{xy} & \sigma_{yy} & \tau_{yz} \\
    \tau_{xz} & \tau_{yz} & \sigma_{zz} \\
  \end{bmatrix}
  \underbrace{
  \begin{pmatrix}
    \cos\theta \\ 0 \\ \sin\theta
  \end{pmatrix}}_{= \underline{m}}
  &=
  \begin{pmatrix}
    \sigma_{xx}\cos\theta + \boldsymbol{\tau_{xz}}\sin\theta \\
    \tau_{xy}\cos\theta + \tau_{yz}\sin\theta \\
    \boldsymbol{\tau_{xz}}\cos\theta + \sigma_{zz}\sin\theta
  \end{pmatrix} = \underline{0}.
\end{align}
Since we can assume that $\sigma_{xx} \approx 0$, it follows that
$\tau_{xz}$ has to be zero. In the case of $\theta = \ang{0}$, even
this assumption is not needed and $\tau_{xz}$ is required to be
exactly zero.

In the $yz$ plane, though, no constraint is placed on $\tau_{xz}$:
\begin{equation}
  \label{eq:boundary-yz}
  \begin{bmatrix}
    \sigma_{xx} & \tau_{xy} & \tau_{xz} \\
    \tau_{xy} & \sigma_{yy} & \tau_{yz} \\
    \tau_{xz} & \tau_{yz} & \sigma_{zz} \\
  \end{bmatrix}
  \begin{pmatrix}
    0 \\ 1 \\ 0
  \end{pmatrix}
  =
  \begin{pmatrix}
    \tau_{xy} \\ \sigma_{yy} \\ \tau_{yz}
  \end{pmatrix}
  = \underline{0}.
\end{equation}
Therefore, the sharp neck at the interface leads to a stress
concentration. Here, the slip mode cannot be treated like plasticity
with a defined shear strength, but slippage is cracklike. As
demonstrated in Section~\ref{sec:results-3d}, a blunt interface can
reduce the stress concentration and disable this cracklike mechanism
in some cases.

\begin{figure}[h!]
  \centering
  \includegraphics{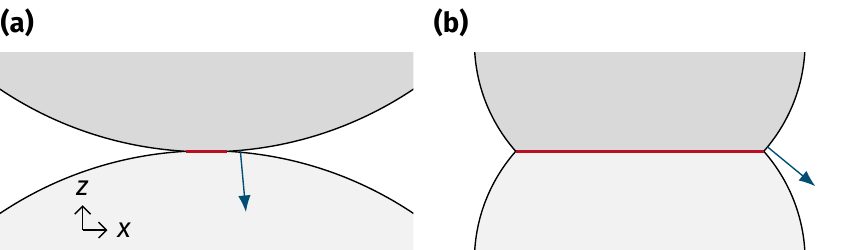}
  \caption{Schematic of the contact geometry assumed (a) in classical
    contact solutions and (b) in our case. The arrows represent the
    surface normal at the edge of contact.}
  \label{fig:contact}
\end{figure}

It should be noted that classical contact solutions---such as
Cattaneo--Mindlin---predict a radially symmetric stress concentration
at the edges of the contact instead of only a lateral one
\cite{Johnson1985}, meaning that stress concentrations should be
present in 2D and 3D geometries. The present case, though, violates the
assumptions of such classical solutions. First, we do not have an
incommensurate, frictional contact of spheres, but a commensurate,
adhesive interface. More importantly, though, the contact radius is
large compared to the radius of the sphere in our
case. Figure~\ref{fig:contact}(a) shows that the surfaces in the
classical solution are approximately parallel at the edge of the
contact area, which means that the free boundary condition used in
Eq.~\ref{eq:boundary-xz} does not exist, while it is present for large
contact radii [Fig.~\ref{fig:contact}(b)]. Additionally, the
nucleation of slip from the sides lateral to the macroscopic sliding
direction has been observed before in rubber--glass contact
\cite{Audry2012} and in atomistic contact simulations of crystalline
lattices \cite{Sharp2016, Sharp2017}. The latter publications ascribe
the asymmetry to nonlinear effects, i.e., that the screw dislocations
nucleated at the lateral sides have a smaller core width and thereby
increase the stress concentration there. We show here that even linear
elasticity can lead to asymmetric slip nucleation, at least for large
contact areas. Nonlinear effects might still be necessary for small
ratios of contact area to sphere radius.

\vfill

\balancecolsandclearpage


\section*{Supplementary material}

# Adhesive wear mechanisms in the presence of weak interfaces: Insights from an amorphous model system

Tobias Brink *and* Jean-François Molinari

*Civil Engineering Institute and Institute of Materials Science and Engineering,
École polytechnique fédérale de Lausanne (EPFL), Station 18, CH-1015 Lausanne, Switzerland*

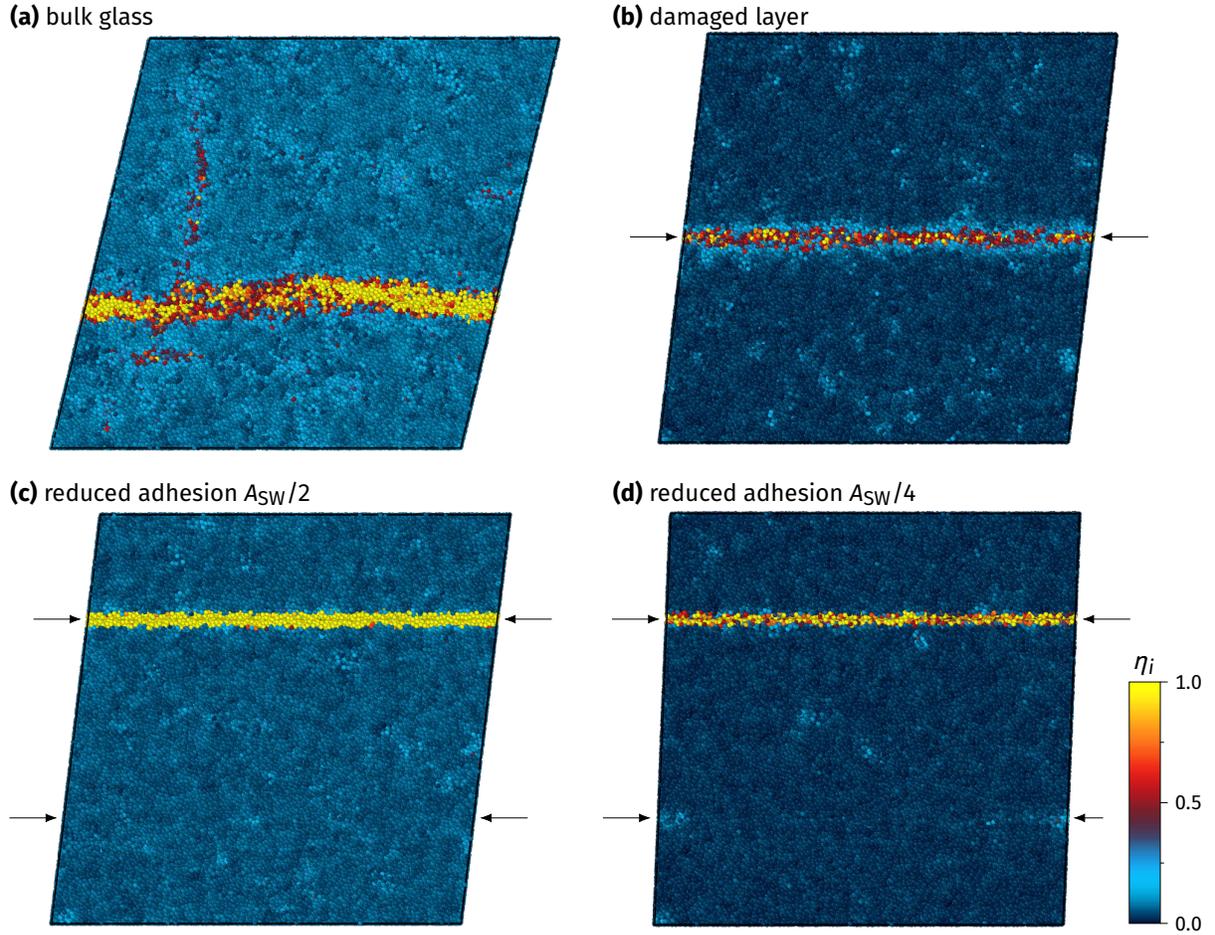

**Fig. S. 1:** Snapshots of bulk shear tests with periodic boundary conditions. All simulation boxes have a size of roughly $20 \times 20 \times 20$ nm$^3$. (a) Homogeneous bulk glass at a macroscopic shear strain of $\gamma = 24\%$, color coded according to the local atomic shear strain $\eta_i$. A single, dominant shear band is visible. (b) Simulation with a damaged layer (arrows) of width 1 nm at $\gamma = 12\%$. The shear band is forced onto the plane of the damaged layer, but some diffuse plasticity can be observed around the shear band. (c)–(d) The simulations with interfaces were modeled by two identical material blocks with a reduced adhesion potential between them. Due to the periodic boundary conditions, this results in two interface planes (arrows). Slip is completely confined to one of these two planes and no significant plasticity is observed next to the interface. The snapshots were taken at shear strains of 12% (c) and 4% (d), respectively.

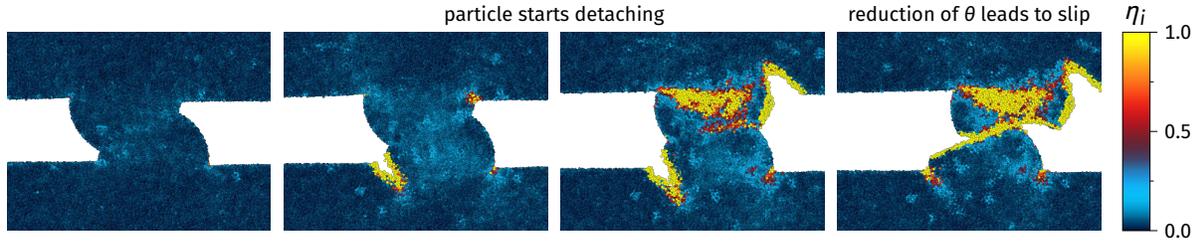

**Fig. S.2:** Slip after partial wear particle detachment due to decreasing inclination. Snapshots from a simulation with interface adhesion strength of $\tilde{\tau}_{\text{adh}} = 0.7\,\tilde{\tau}_b$, $d = 10$ nm, $p_N = 0.1\,\tilde{\tau}_b$, and initial $\theta = 30°$. At the onset of fracture, $\theta = 25°$. As soon as the interface rotates to $\theta \approx 15°$, slip sets in.

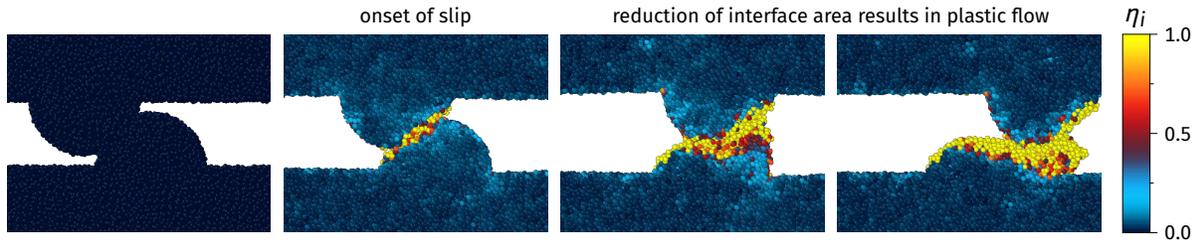

**Fig. S.3:** Plastification during slip due to increasing normal pressure. Snapshots from a simulation with interface adhesion strength of $\tilde{\tau}_{\text{adh}} = 0.15\,\tilde{\tau}_b$, $d = 5$ nm, $p_N = 0.1\,\tilde{\tau}_b$, and initial $\theta = 45°$. The junction area reduces after the onset of slip, leading to an increase of $p_N$ and plastic flow.

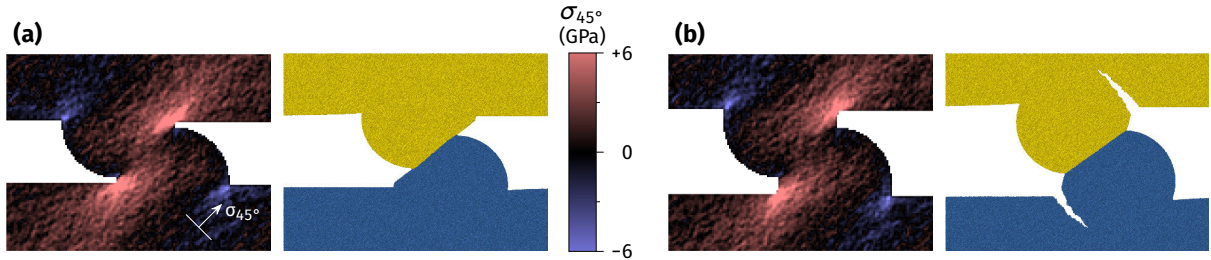

**Fig. S.4:** Sensitivity of the interface to stress concentrations. Simulations with interface adhesion strength of $\tilde{\tau}_{\text{adh}} = 0.15\,\tilde{\tau}_b$, $d = 30$ nm, $p_N = 0.1\,\tilde{\tau}_b$, and initial $\theta = 45°$. (a) If the inclination $\theta$ is high, its edge can come close to the stress concentration at the corner of the asperity. This seems to lead to the nucleation of a slip front even in cases where our model predicts particle detachment. (b) By putting the asperities on small "sockets" (height $\approx d/10$), the stress concentrator is moved away from the junction interface and a wear particle detaches, as expected. The latter procedure was performed for the simulations with initial $\theta \geq 45°$ in Fig. 5(c) and (d).

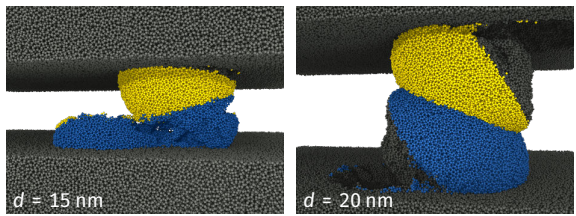

**Fig. S.5:** Snapshots of simulations where the interface is stronger than the bulk of the asperities. Simulations were prepared by randomly removing 5% of atoms in the asperities, keeping the 2-nm-wide interface intact. This reduces the bulk shear strength to 6.9 GPa and the shear modulus to roughly 45 GPa, resulting in $d^* \approx 17$ nm. The simulations displayed on the left (initial $\theta = 30°$) confirm this value.

2

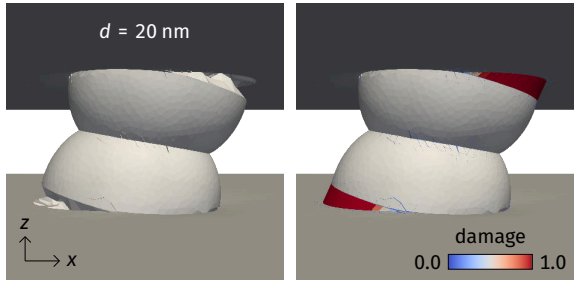

**Fig. S.6:** FEM simulation of a contact junction without weakened interface. The left image shows only the solid, while the right image includes the cohesive elements. The cracks appear, as expected, at the bottom of the asperities and lead to the detachment of a wear particle.

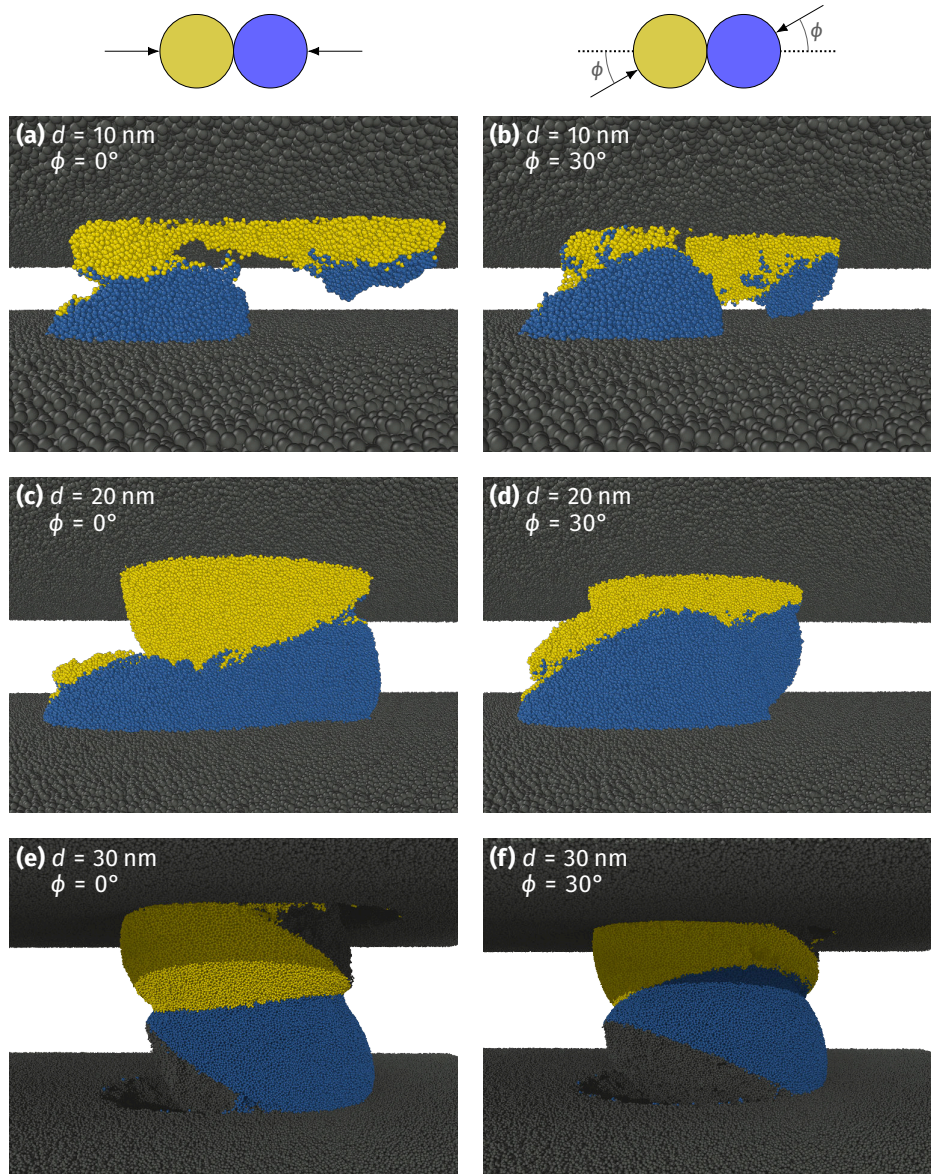

**Fig. S.7:** Simulations with a sliding direction deviating from the $xz$ plane. It can be seen here—as expected from the derivation in Appendix B—that the mechanism is unchanged for a rotation of the sliding direction by $\phi = 30°$: Asperities with $d \leq 20$ nm plastify (a)–(d), while asperities with $d = 30$ nm form wear particles (e)–(f), independent of $\phi$.

3


\end{document}